# Designing Dual-Band Absorbers by Graphene/Metallic Metasurfaces

Saeedeh Barzegar-Parizi, Amin Khavasi

*Abstract*-- This article presents a novel approach for designing dual-band absorbers based on graphene and metallic metasurfaces for terahertz and mid-infrared regimes, respectively. The absorbers are composed of a two-dimensional (2D) array of square patches deposited on a dielectric film terminated by a metal plate. Using an analytical circuit model, we obtain closed-form relations for different parameters of the structure to achieve the dual-band absorber. Two absorption bands with obtained absorptivity of 98% at 0.53 and 1.53 THz for the graphene-based structure and 7 and 25 THz for the metallic-based case are achieved. We demonstrate that the graphene-based absorber remains as the dual-band for a wide range of the Fermi level. Furthermore, the recommended dual-band absorbers are insensitive in terms of polarization and remain within various incident angles. The most important advantage of this device is its simplicity compared to previously reported structures.

*Index Terms*- Dual band, Absorber, THz frequencies, Mid-infrared, Patch array, Circuit model.

## I. INTRODUCTION

Metamaterial-based absorbers have received considerable attention in the optics and microwave and have been under investigation for many years [1-6]. Metamaterial absorbers have demonstrated promising applications in solar energy harvesting [7,8], refractive index sensors [9], microbolometers [10], thermal imaging [11], thermal IR emitters [12], bio-sensing, and etc. The absorbers studied in previous works can be classified into three categories: wideband, multiband, and narrow absorbers. To achieve wideband/multiband absorbers, various mechanisms of broadband/multiband absorption have been recommended by a combination of two or more resonators with different sizes together [13–16] or using the stacked structures composed of the multiple layers of resonators with diverse geometric dimensions parted by dielectric layers [17-20] or asymmetrical-shaped resonators [21,22]. However, these methods complicate the fabrication process and result in high costs. Therefore, designing simple structures with single step lithography is of great importance [23-30].
Recently, simple structures composed of the patterned arrays of graphene and metals placed on a dielectric spacer above a reflecting surface have been employed for achieving broadband absorption [27-30]. Using the analytical circuit model for absorber and impedance matching concept, a normalized absorption bandwidth of 100 % has been achieved.

In this paper, we propose a novel approach for designing absorbers for dual-band applications, which are insensitive to polarization and omnidirectional for low terahertz and mid-infrared regimes. We use arrays of graphene and metal patches to develop dual-band absorbers in low terahertz and mid-IR regimes, respectively. The proposed absorber is composed of a simple structure: a 2D periodic array of graphene/metal patches placed on a quarter wavelength dielectric film above a reflecting surface. Furthermore, the closed-form associations have been introduced for the geometrical parameters of the device and the properties of the material in this paper.
A circuit model for the device is recommended where the admittance of arrays of graphene/metal patches has been presented in [31]. Thanks to the analytical circuit model, we present a rigorous approach for achieving high absorption level at two frequencies. The impedance matching concept is employed at two frequencies to obtain high absorption where the input impedance of the circuit approaches that of free space. We design high-performance dual-band absorbers in low terahertz and mid-infrared regimes using the proposed approach. Then, full-wave simulations are applied to confirm the design method. It should be noted that the recommended absorber is polarization insensitive for normal incident EM waves because its structure is symmetric. Besides, both dual-band and the near-perfect absorption effectiveness are not influenced significantly even at high angles of incidence for both TE and TM polarizations.

The paper is structured as follows: In Sec. II, the circuit model of the proposed device is introduced, and then a design procedure is recommended that results in the dual-band absorber. The limitations of the recommended design are discussed in Sec. III. Several numerical examples is given and compared with the results of the commercial software HFSS in Sec. IV.

## II. DESIGN METHOD OF DUAL-BAND ABSORBER

The Salisbury absorbers are commonly made of three layers: a top metamaterial including patterned arrays of one-dimensional or two-dimensional subwavelength elements, a middle dielectric layer, and a reflective metallic lowest layer [32]. It is believed that absorption is done perfectly if the structure and the incident medium are matched in terms of the impedance [29].
For the development of EM absorber in the low terahertz regime, many have used graphene-based patterned arrays [33-37]. In this section, the graphene patch array positioned on a dielectric film over a thick enough metallic film is discussed in this section as displayed in Fig.1(a) to develop the dual-band absorber. In comparison to previous works, we use a simple structure instead of merging two or more resonators

S. Barzegar-Parizi is with the Electrical Engineering Department, Sirjan University of Technology, Sirjan, Iran. (barzegarparizi@sirjantech.ac.ir)
A. Khavsi is with the Electrical Engineering Department, Sharif University of Technology, Tehran, Iran.

with diverse sizes [13–16] or stacking several layers of the resonators with diverse geometric sizes [17-20] or asymmetrically-forms [21,22].

The array involves squares with a cross-section of $w \times w$ set along the $x$- and $y$-directions with period $D$. The width and the refractive index of the dielectric film are $h$ and $n_s$, correspondingly.

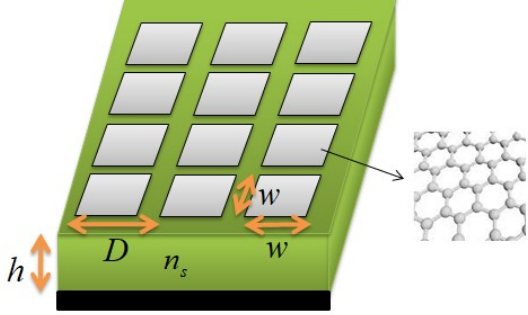

Fig. 1. Structure of an EM absorber composed of a graphene patch array positioned on top of a dielectric film completed by a rear reflector.

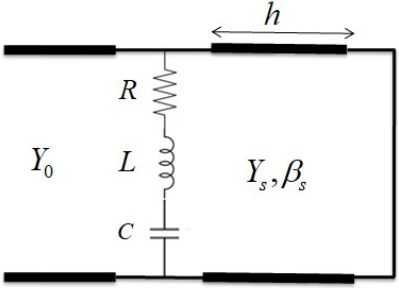

Fig. 2. The equivalent circuit model for the proposed absorber

The equivalent circuit model for the proposed structure is demonstrated in Fig.2. In this circuit, the patch array illuminated by a normally incident plane wave can be modeled by a surface admittance [31]

$$Y_{sur} = \frac{S^2}{D^2 K}(\sigma_g^{-1} + \frac{q}{j\omega\varepsilon_{eff}})^{-1} \quad (1)$$

Where, $S = 1.015\, w^2$, $K = 1.1634\, w^2$ and $q$ is the eigenvalue is the eigenvalue matching the basic resonant mode of the array whose value versus $w/D$ has been calculated by a variational approach in [31]. In this equation $\varepsilon_{eff} = \varepsilon_0(1+n_s^2)/2$, and $\sigma_g$ which is the surface conductivity of graphene is given by:

$$\sigma_g = \frac{2e^2 k_B T}{\pi \hbar^2} \frac{j}{-\omega + j\tau^{-1}} \ln[2\cosh(E_F/2k_B T)]$$
$$- \frac{je^2}{4\pi \hbar} \ln\left[\frac{2E_F - \hbar(\omega - j\tau^{-1})}{2E_F + \hbar(\omega - j\tau^{-1})}\right] \quad (2)$$

Where $e$ is the electron charge, $E_F$ denotes the Fermi level, $\hbar$ refers to the reduced Plank constant, $k_B$ represents the Boltzmann constant, $\omega = 2\pi f$ shows the angular frequency, $T = 300$ K is the temperature and $\tau$ is the relaxation time. At appropriately low frequencies (where the inter-band term of conductivity, the second term in the above equation, is insignificant) and for $E_F \gg k_B T$, $\sigma_g$ will be of the Drude form:

$$\sigma_g = \frac{e^2 E_F \tau}{\pi \hbar^2} \frac{1}{1+j\omega\tau} \quad (3)$$

According to (1) and (3), the surface admittance appears as a R-L-C series branch as illustrated in Fig. 2. The values of R, L, and C are obtained by the following relations:

$$R = \frac{D^2 K}{S^2} \frac{\pi \hbar^2}{e^2 E_F \tau},\ L = \tau R,\ C = \frac{S^2}{D^2 K} \frac{\varepsilon_{eff}}{q} \quad (4)$$

In the equivalent circuit model (Fig. 2), $\beta_s = k_0 n_s$ and $Y_s = n_s/\eta_0$ are the propagation constant and the admittance of transmission line matching the dielectric slab respectively, where $\eta_0 = 120\pi$ is the free-space impedance and $k_0 = \omega/c$ ($c$ is the speed of light) is the wavenumber of free space. The metallic back reflector can be approximately considered as a short circuit [29]. It is accurate enough for terahertz and mid-infrared regimes as long as the metallic back plate is sufficiently thick [29]. Now we have the input admittance of the device:

$$Y_{in} = Y_g - jY_s \cot(\beta_s h) \quad (5)$$

To develop the dual-band absorber, we consider a central frequency defined as $f_0$ in which the transmission line equivalent to the dielectric slab acts as a quarter wavelength line by setting $\beta_s h = \pi/2$. Hence, the input admittance of the absorber will be $Y_g$. Then, the thickness of the dielectric slab is obtained as:

$$\beta_s h = \frac{\pi}{2} \Rightarrow h = \frac{c}{4 n_s f_0} \quad (6)$$

We also consider:

$$LC = \frac{1}{\omega_0^2} \quad (7)$$

Hence, at central frequency, the input admittance of the device is represented as $Y_{in} = R$. Now, we should impose the conditions to achieve high absorption at two frequencies $\omega_1$ and $\omega_2$. Accordingly, the conditions are written as:

$$\mathrm{Im}(Y_{in})\big|_{\omega=\omega_1} = 0 \quad (8.a)$$
$$\mathrm{Re}(Y_{in})\big|_{\omega=\omega_1} = \alpha_1/\eta_0 \quad (8.b)$$

and

$$\mathrm{Im}(Y_{in})\big|_{\omega=\omega_2} = 0 \quad (9.a)$$
$$\mathrm{Re}(Y_{in})\big|_{\omega=\omega_2} = \alpha_2/\eta_0 \quad (9.b)$$

Where the values of $\alpha_1$ and $\alpha_2$ should be chosen such that the absorption values are high (for example above 90%). To satisfy this condition, we should have $0.52 < \alpha_{1,2} < 1.92$. Applying (8.b) and (9.b) leads to:

$$\tau^2 \omega_0^2 = \frac{\alpha_2 - \alpha_1}{\left(\alpha_1 (\frac{\omega_1}{\omega_0})^2 \left(1-(\frac{\omega_0}{\omega_1})^2\right)^2 - \alpha_2 (\frac{\omega_2}{\omega_0})^2 \left(1-(\frac{\omega_0}{\omega_2})^2\right)^2\right)} \quad (10)$$

On the other hand, we obtain from (8)-(9):

$$\alpha_i = \frac{-n_s \cot(\beta_s h)|_{\omega=\omega_i}}{\tau \omega_i \left(1-(\frac{\omega_0}{\omega_i})^2\right)} \quad (11)$$

with $i = 1$ and 2.

After straightforward mathematical manipulations, (10) and (11) reduce to:

$$\frac{n_s^2 \cot^2\left(\frac{\pi \omega_2}{2\omega_0}\right)}{\alpha_2} - \frac{n_s^2 \cot^2\left(\frac{\pi \omega_1}{2\omega_0}\right)}{\alpha_1} = \alpha_1 - \alpha_2$$

$$\alpha_2 = \alpha_1 \frac{\omega_1 \left(1-(\frac{\omega_0}{\omega_1})^2\right) \cot(\frac{\pi \omega_2}{2\omega_0})}{\omega_2 \left(1-(\frac{\omega_0}{\omega_2})^2\right) \cot(\frac{\pi \omega_1}{2\omega_0})} \quad (12)$$

Choosing the first normalized frequency ($\omega_1 / \omega_0$) and using (12), one can obtain the proper values for the second normalized frequency ($\omega_2 / \omega_0$) and the parameters of $\alpha_{1,2}$ for given values of $n_s$. However, the condition $0.52 < \alpha_{1,2} < 1.92$ should always be applied.

After calculation, the values of $\alpha_{1,2}$ and $\omega_{1,2}/\omega_0$, the multiplication of the central frequency, and the relaxation time of graphene can be extracted by Eq. (11). The relaxation time can be tuned by the Fermi level $E_f$ through the relation $\tau = \frac{E_f \mu}{e v_f^2}$, in which $\mu$ is the electron mobility ranging from about 0.03 m²/Vs to 6 m²/Vs depending on the fabrication process [38-41] and $v_f = 10^6$ m/s is the Fermi velocity. Thus, by computing the relaxation time, the Fermi level is computed by:

$$E_f = \frac{\tau e v_f^2}{\mu} \quad (13)$$

In addition, using (8.b) and (9.b), we have

$$R \left(1 + \tau^2 \omega_i^2 \left(1-(\frac{\omega_0}{\omega_i})^2\right)^2\right) = \frac{\eta_0}{\alpha_i} \quad (14)$$

Considering

$$R = \frac{\eta_0}{\alpha_0} \quad (15)$$

the parameter of $\alpha_0$ is computed as:

$$\alpha_0 = \alpha_i \left(1 + \tau^2 \omega_i^2 \left(1-(\frac{\omega_0}{\omega_i})^2\right)^2\right) \quad (16)$$

Indeed, $\alpha_0$ specifies the absorption value at the central frequency, where the input admittance appears as $Y_{in} = R$; hence, the reflection becomes as:

$$\Gamma = \left(\frac{1-\alpha_0}{1+\alpha_0}\right) \quad (17)$$

According to (16), $\alpha_0$ cannot be chosen arbitrarily and depends on the properties of absorption at the frequency of two sides. At the end, the geometrical parameters of the structure are computed by (15) and (7) as:

$$\frac{D^2 K}{S^2} \frac{\pi \hbar^2}{e^2 E_F \tau} = \frac{\eta_0}{\alpha_0} \Rightarrow \frac{w}{D} = \sqrt{\frac{\alpha_0 \pi \hbar^2}{0.868 \eta_0 e^2 E_f \tau}} \quad (18.a)$$

$$LC = \frac{1}{\omega_0^2} \Rightarrow w = \frac{e^2 E_f}{\pi \hbar^2 \varepsilon_{eff}} \frac{r}{\omega_0^2} \quad (18.b)$$

with $r = wq$.

### III. LIMITATIONS

In this section, we survey the limitations on the design algorithm presented in the previous section. The first limitations terms from Eq.(12) as $\alpha_{1,2}$ must satisfy the condition $0.52 < \alpha_{1,2} < 1.92$ for achieving the absorption conditions above 90%. It is clear that for a specified value of the refractive index and the normalized frequency of the first band ($\omega_1/\omega_0$), this condition causes limitation on the selection of $\omega_2/\omega_0$. One can simply extract the expressions for $\alpha_1$ and $\alpha_2$ as a function of $\omega_2/\omega_0$ from (12). Fig. 3 exhibits the changes of $\alpha_{1,2}$ as a function of $\omega_2/\omega_0$ for two different values of $\omega_1/\omega_0$. The refractive index is considered as $n_s = 3.13$ corresponding to Al$_2$O$_3$. As observed, the second-band normalized frequency can be obtained with respect to the limitation on the parameters of $\alpha_i$. To achieve near-one absorption conditions, it suffices to select parameters of $\alpha_i$ close to one.

As mentioned previously, the parameter of $\alpha_0$, which specifies the absorption characteristics in the middle of the two bands, depends on the values of the parameters of $\alpha_i$. Fig. 4 reveals the changes of $\alpha_0$ as function of $\omega_2/\omega_0$ for the refractive index of $n_s = 3.13$.

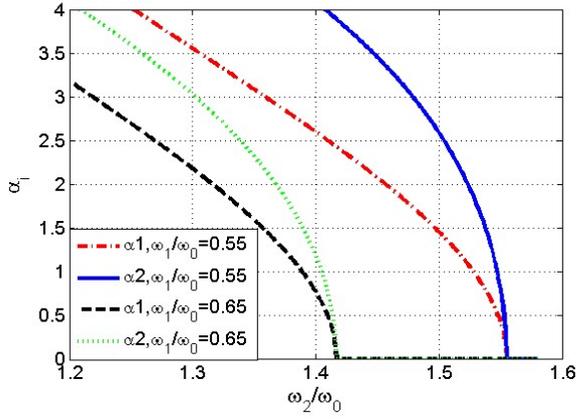

Fig. 3. The parameters of $\alpha_1$ and $\alpha_2$ as a function of $\omega_2/\omega_0$ for $n_s = 3.13$

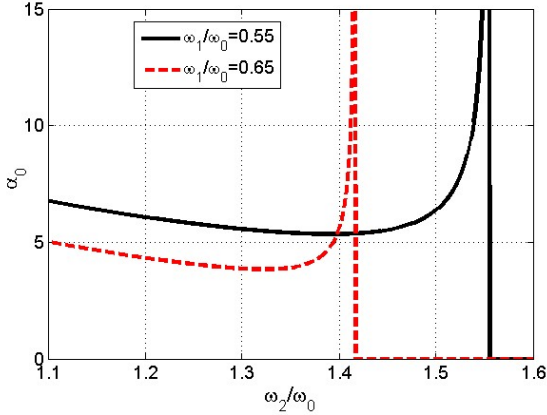

Fig. 4. The parameters of $\alpha_0$ as function of $\omega_2/\omega_0$ for $n_s = 3.13$.

Another limitation comes from Eq. (18.a) since $w/D$ must be smaller than unity ($w/D < 0.9$). This limitation influences the selection of the electron mobility ($\mu$) and the central frequency $\omega_0$ as:

$$\mu \omega_0^2 < \frac{0.7031 \eta_0 e^3 v_f^2 n_s^2 \alpha_2^2}{\pi \hbar^2 \alpha_0} \left( \frac{\cot^2\left(\frac{\pi \omega_2}{2\omega_0}\right)}{(\omega_2/\omega_0)^2 \left(1 - \left(\frac{\omega_0}{\omega_2}\right)^2\right)^2} \right)$$

(19)

The structure is of subwavelength; therefore, the period length must be smaller than the lowest wavelength $D < \lambda_{\min}$, where $\lambda_{\min} = c/n_s f_{\max}$ ($f_{\max}$ should be considered larger than the second band where we consider $f_{\max} = \beta f_0$ with $\beta > 1.5$). According to (18.b), this limitation appears as:

$$\mu \omega_0^2 > \frac{0.55 \beta e^3 v_f^2 n_s^2 \alpha_2}{\pi^2 \hbar^2 \varepsilon_{eff} c} \left( \frac{\cot\left(\frac{\pi \omega_2}{2\omega_0}\right)}{(\omega_2/\omega_0)\left(1 - \left(\frac{\omega_0}{\omega_2}\right)^2\right)} \right) \quad (20)$$

Fig. 5 shows the limitations due to (19) and (20) as a function of $\omega_2/\omega_0$ for the refractive index of $n_s = 3.13$. The proper values of $\mu f_0^2$ are limited to the region below the blue color curve and above the red color curve. Accordingly, by choosing the normalized frequency of the second band from Fig. 3 and taking into account the range specified for $\mu$ [38-41], the values for the center frequency and $\mu$ are obtained from Fig. 5.

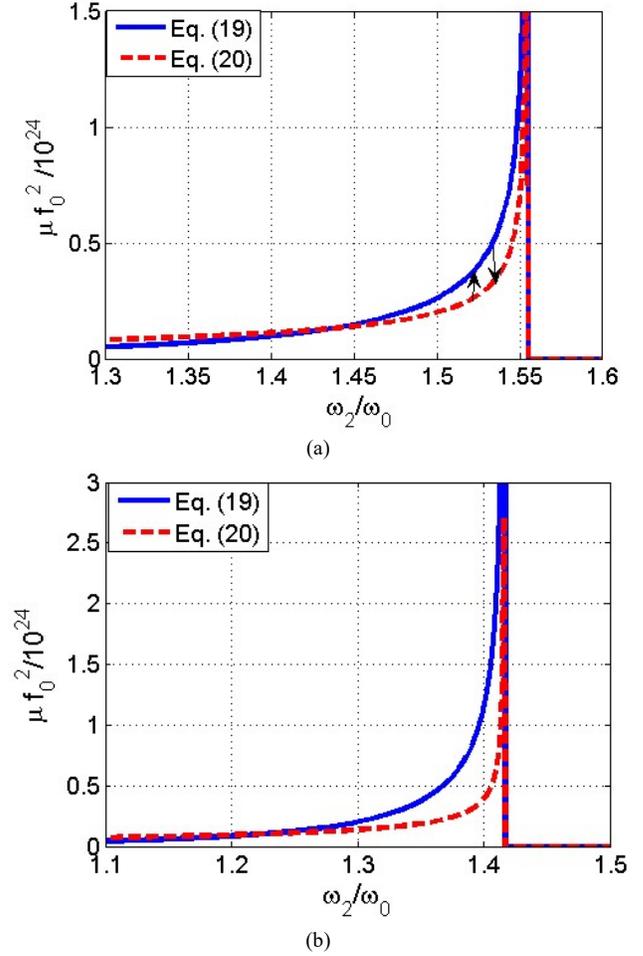

Fig. 5 Eqs. (19) and (20) as a function of $\omega_2/\omega_0$ for the refractive index of $n_s = 3.13$ with the normalized frequency of the first band (a) $\omega_1/\omega_0 = 0.55$ and (b) $\omega_1/\omega_0 = 0.65$

In the next section, the dual-band absorbers would be designed by using the aforementioned principles and relations.

IV. NUMERICAL RESULTS

In this section, the proposed design procedure is verified through some numerical examples.
For electrostatic biasing of graphene, we use two polysilicon DC gating sheets of distance t = 50 nm as shown in Fig. 6. The bias voltage is applied to polysilicon sheets to modulate the Fermi level of the graphene patches. The graphene layer has been deposited on the top polysilicon sheet. To avoid the effect of polysilicon sheets on the electromagnetic response of the device, the thicknesses of the polysilicon sheets is considered very small ($t_p$ = 100 nm).

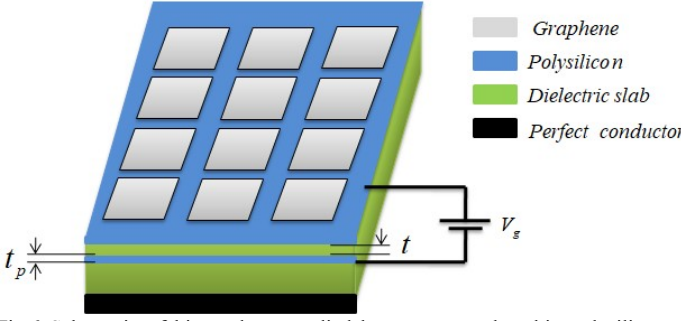

Fig.6 Schematic of bias voltage applied between two ultra-thin polysilicon sheets in order to modulate the Fermi level of the graphene patches. t=50 nm is the distance between two polysilicon sheets, and tp=100 nm is the thicknesses of the polysilicon sheets. These values of t and tp are chosen according to [43].

As the first example, the features of graphene and the geometrical factors of the structure for the design of a dual-band absorber at THz frequencies. Suppose that the refractive index of the dielectric film is $n_s = 3.13$ and the frequency of the first band and second band are $f_1 = 0.55$ THz and $f_2 = 1.54$ THz. Hence, the values of coefficients $\alpha_{1,2}$ are extracted as $\alpha_1 = 0.735$ and $\alpha_2 = 1.39$ from Fig.3. The central frequency is considered as $f_0 = 1$ THz. The electron mobility is $\mu = 0.6$ m$^2$/Vs from Fig. 5a and the desired relaxation time will be $\tau = 4.56 \times 10^{-13}$ S from (10); thus using (13), we have $E_f = 0.76$ eV. The thickness of the dielectric spacer is given by Eq. (6) as $h = 23.96 \,\mu m$. Finally, from Eqs. (18.a) and (8.b), the patch width and the period are extracted as $w = 28.27 \,\mu m$ and $D = 32 \,\mu m$ respectively with $\alpha_0 = 10.4$ obtained from Fig.4. Fig. 7 demonstrates the absorption spectra for the designed device. The result obtained the analytical model is compared with the result obtained by full-wave simulations (HFSS), which are in a good agreement. For full-wave simulations conducted by HFSS, graphene is modeled by a layer of thickness $\Delta = 1 nm$ whose permittivity is $\varepsilon_g = \varepsilon_0 - j\sigma_g /(\Delta\omega)$ [24,30]. The simulation is performed on a single period with periodic boundary settings. As observed, a dual-band absorber with absorption above 98% can be developed.

Fig.8 shows the loss distribution on a graphene patch at two resonant frequencies to better understand the absorption mechanism. It is worth mentioning that the power loss density distributions are similar for both absorption peak frequencies. Since the fundamental resonant mode (first surface Plasmon polariton) of graphene patch is responsible for both resonant frequencies. It is in contrast to previous works [23-25] where the absorption is mainly due to the excitation of the fundamental and second-order SPRs at the first and the second absorption peaks, respectively.

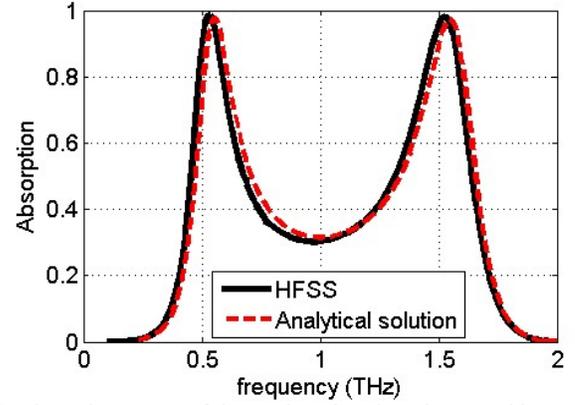

Fig. 7. Absorption spectra of the graphene-based patch array with properties $w = 28.27 \,\mu m$ and $D = 32 \,\mu m$. The properties of graphene appear as $\tau = 4.56 \times 10^{-13}$ s and $E_f = 0.76$ eV.

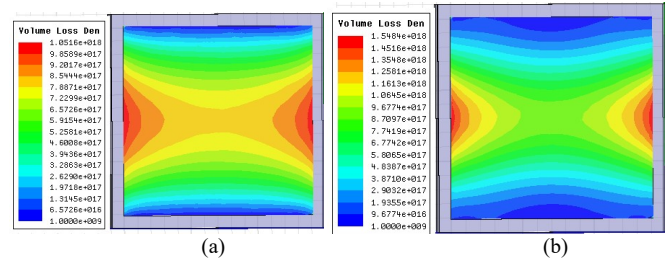

(a)          (b)

Fig.8 The loss density distributions on the graphene patch (a) f$_1$ = 0.53 THz and (b) f$_2$ = 1.53 THz.

Fig. 9 reveals the input admittance as a function of frequency for the absorber. As observed, the imaginary part of the surface admittance is zero at three frequencies of $f_0$ and $f_{1,2}$. The real part of admittance is very high at the central frequency (10 times larger than the free space admittance) and closely matches free space admittance at frequencies of both sides. Therefore, the condition is prepared for impedance matching at two bands and based on achieving a dual-band absorber as shown in Fig. 7.

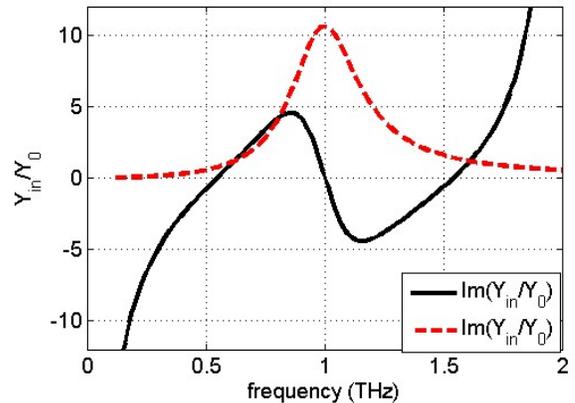

Fig.9 The normalized input admittance of graphene-based absorber with parameters presented in Fig.7.

Now, we investigate the omnidirectional characteristic for the proposed absorber. Figs. 10 (a) and (b) show the absorption spectra as a function of incident angle and frequency for graphene patches for TE and TM polarization, respectively. The two absorption peaks can be observed for large incident angles indicating the omnidirectional characteristic of the proposed absorber.

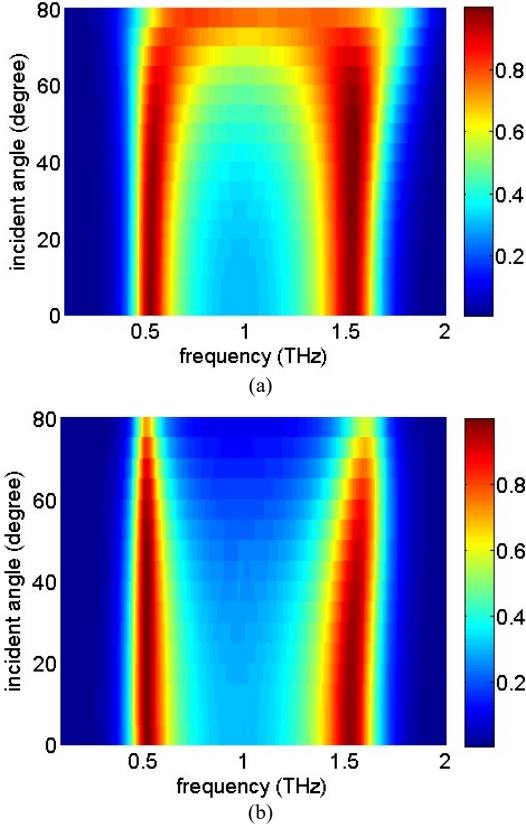

Fig. 10 Absorption spectra of the graphene-based absorber as a function of incident angle and frequency for (a) TM and (b) TE polarizations.

In the next step, we survey the Fermi level changes on the absorption spectra. The Fermi level of graphene can be regulated in a varied range by altering the carrier mobility via chemical doping method or electrical gating. Fig. 11 shows the absorption spectra as a function of the Fermi level of the graphene and the frequency for the designed absorber. It is demonstrated that for the wide range of the Fermi level, two absorption bands can be achieved. It is worth to describe that one can design the dual-band absorber at higher THz frequencies with the new design of the geometrical parameters and adjusting the graphene properties.

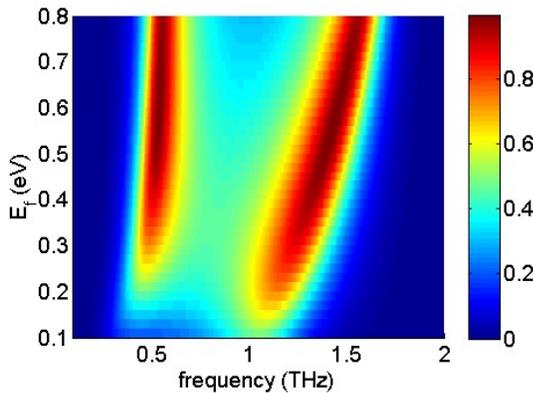

Fig. 11 Absorption spectra of the graphene-based proposed absorber as function of the Fermi level of graphene and frequency.

Next, we design a dual-band absorber in the mid-infrared regime. To achieve this goal, one can use the metallic-based patterned array. In this section, the approach suggested in the previous section was applied to develop EM absorber in the mid-infrared regime, where the graphene is substituted by a metal with the conductivity $\sigma(\omega)$. If the wavelength of the incident plane wave approached mid-infrared and optical ranges, the metals display plasmonic resonance instead of the typical conductivity. So, the conductivity of the metallic patterns is explained as the Drude model:

$$\sigma(\omega) = \frac{\sigma_0}{1 + j\omega\tau} \tag{21}$$

Where $\omega$ is the angular frequency, $\tau$ is the relaxation time, and $\sigma_0$ shows the dc conductivity of the metal computed as $\sigma_0 = \frac{\varepsilon_0 \omega_p^2}{\gamma}$. Here, $\omega_p$ stands for the plasma frequency of the free electron gas and $\gamma$ denotes the characteristic collision frequency defined as $\gamma = 1/\tau$. Table II presented in [30] exhibits the values of $\omega_p$ and $\gamma$ for different metals in the IR regime [43]. The thickness of the bottom reflector layer should have been considered sufficiently thick. In this way, the PEC approximation would be reasonable. The surface admittance of the metallic patch array appears as Eq. (1), where $\sigma_g(\omega)$ replaces $\sigma(\omega)\Delta$ ($\Delta$ represents the thickness of metallic patterns). Therefore, R, L, and C of the circuit model of Fig. 2 are obtained as:

$$R = \frac{D^2 K}{\sigma_0 \Delta S^2}, \quad L = \tau R, \quad C = \frac{S^2}{D^2 K} \frac{\varepsilon_{eff}}{q} \tag{22}$$

Similar to the previous procedure to achieve dual-band absorber, the relaxation time of metal is computed from (10). The most proper metal corresponding to this relaxation time can be considered based on Table I presented in [30]. The geometry parameters of the structure are computed as:

$$h = \frac{c}{4 n_s f_0} \tag{23.a}$$

$$\frac{w}{D} = \sqrt{\frac{\alpha_0}{0.868 \eta_0 \sigma_0 \Delta}} \tag{23.b}$$

$$w = \frac{\sigma_0 \Delta r}{\tau \varepsilon_{eff} \omega_0^2} \tag{23.c}$$

Next, the metal features and the geometry factors of the structure for the design of a dual-band absorber with the frequency of the first band $f_1 = 7$ THz and the second band $f_2 = 25$ THz are extracted. The dielectric film is considered $Al_2O_3$. The values of coefficients $\alpha_{1,2}$ are selected as $\alpha_1 = 0.59$ and $\alpha_2 = 1.47$ (the central frequency is $f_0 = 15$ THz). Hence, the relaxation time of the metal appears as $\tau = 3.8 \times 10^{-14}$ s leading to $\gamma/(2\pi c) = 1.4 \times 10^4$ $(m^{-1})$, which corresponds to Ag metal (see [30]) with $\sigma_0 = \frac{\varepsilon_0 \omega_p^2}{\gamma} = 6 \times 10^7$ $(\Omega^{-1} m^{-1})$. Using (23), the geometrical parameters appear as $\Delta = 1.3 nm$, $h = 1.6 \mu m$, $w = 2.92 \mu m$, and $D = 3.23 \mu m$ for $\alpha_0 = 21$. Fig. 12 displays the absorption spectra for the designed device. As can be observed, two absorption bands have been obtained with absorptivity of 98% at 7 and 25 THz.

Figs. 13 (a) and (b) reveal the absorption spectra as a function of azimuth incident angle and frequency for TE and TM polarization, correspondingly. According to these figures, the recommended absorber performs well within a wide range of incident angles for both TM and TE polarizations. Furthermore, for the incident angle below $60^0$, both polarizations overlap significantly and hence the absorber is not polarization dependent within this range. Accordingly, the absorber exhibits the ultrathin, dual-band absorption, insensitivity to polarizations, and extensive incident angles characteristics.

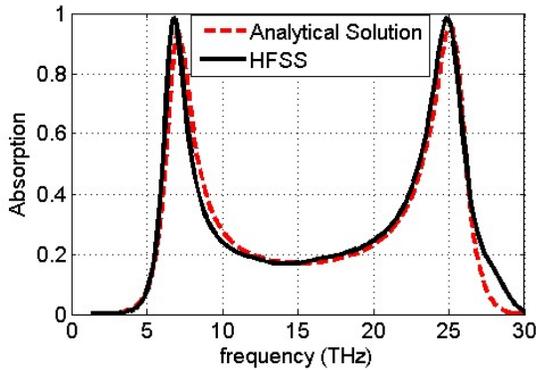

Fig. 12 Absorption spectra of the metallic-based patch array with parameters $w = 2.92 \mu m$, $D = 3.23 \mu m$, and $h = 1.6 \mu m$ where the patches are constructed from Ag metal.

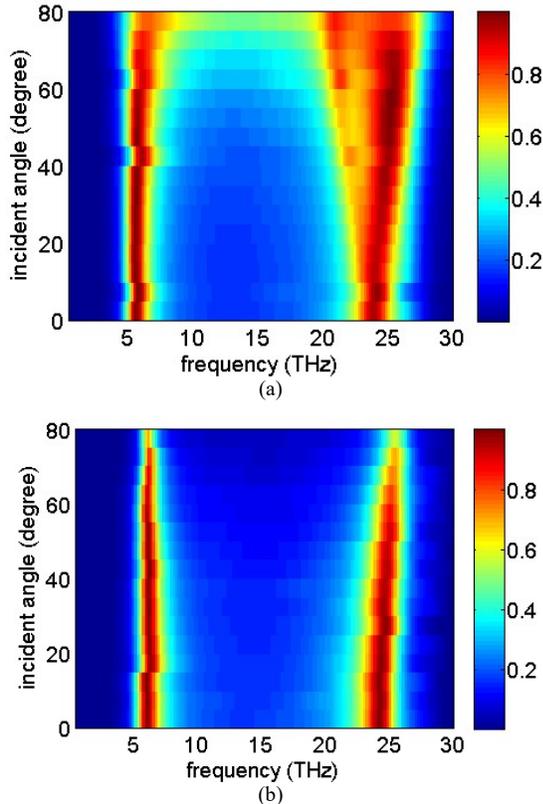

Fig. 13 Absorption spectra of the metallic-based absorber as a function of incident angle and frequency for (a) TM and (b) TE polarizations.

## V. Conclusion

In the present study, dual-band absorbers were examined for mid-IR and low terahertz regimes. The impedance matching settings including regulating the real part to approach the free space admittance and the imaginary part to be zero were set at two frequencies to develop the dual-band absorber. This method resulted in closed-form relations for the geometry of the structure and the characteristics of the applied material. According to the results of simulation and analytical circuit model, the recommended absorber can operate with an absorption value of 98% at 0.53 and 1.53 THz for graphene-based structure and 7 and 25 THz for the metallic-based case. Further, these devices are insensitive to polarization and omnidirectional.

**Saeedeh Barzegar-Parizi** received the B.Sc. degree from the Iran University of Science and Technology, Tehran, Iran, in 2008, and the M.Sc. and Ph.D. degrees from the Sharif University of Technology, Tehran, in 2010 and 2015, respectively, all in electrical engineering. She has been with the Department of Electrical Engineering, Sirjan University of Technology, where she is currently an Assistant Professor. Her research interests include the numerical solving of periodic structures, such as periodic rough surfaces and artificial structures.

**Amin Khavasi** was born in Zanjan, Iran, in 1984. He received the B.Sc., M.Sc., and Ph.D. degrees from the Sharif University of Technology, Tehran, Iran, in 2006, 2008, and 2012, respectively, all in electrical engineering. He has been with the Department of Electrical Engineering, Sharif University of Technology, where he is currently an Assistant Professor. His current research interests include photonics, the circuit modeling of photonic structures, and computational electromagnetic.